# Synchronized Fractionation and Phase Separation in Binary Colloids

Lian Dan Yao,[a] Hong Yu Chen,[a] Yan Shi,[a] Ying Liang,[a] Tian Hui Zhang[a,b*]

[a] *Center for Soft Condensed Matter Physics and Interdisciplinary Research, Soochow University , Suzhou, 215006, P. R. China*

[b] *School of Physical Science and Technology, Soochow University, Suzhou, 215006, P. R. China*

**Corresponding Author**:   Tian Hui Zhang:   zhangtianhui@suda.edu.cn

**Abstract**

Fractionation is necessary for self-assembly in multicomponent mixtures. Here, reversible fractionation and crystallization are realized and studied in a two-dimensional binary colloids which is supersaturated by enhancing the attraction between colloidal particles. As a deep and fast supersaturation results in gels with a uniform distribution of binary particles, a gradual quasistatic supersaturating process leads to a two-step crystallization in which small particles and large particles are fractionated as coexisting crystal and liquid phases respectively. Fractionation occurs as well in the quasistatic melting of gel. We show that the synchronized fractionation and phase separation arises from the competition between the size-dependent repulsion and the tunable attraction. The results in this study demonstrate a robust mechanism of fractionation via phase separation, and have important implication in understanding the reversible formation of membraneless organelles in living cells.

# Introduction

Crystallization or self-assembly in multicomponent mixtures is ubiquitous in physics and biology. It is widely involved in the growth of alloys,[1] the fabrication of complex functional structures,[2] and the formation of membraneless organelles in living cells.[3,4] Since colloidal systems exhibit similar phase behaviors as that observed in atomic systems, they have been widely employed as model systems to study the fundamental principles of self-assembly.[5] These studies have produced rich results and greatly extended our understanding on the dynamics of nucleation and crystal growth.[6] However, different from atomic systems, colloidal particles are generally polydisperse in size. In hard sphere systems, it was found that the polydispersity contained by a crystal cannot be higher than 7%,[7,8] and the presence of polydispersity can dramatically delay[9,10] or hinder the nucleation of crystal.[11] Although studies have shown that introducing long-range soft interactions can significantly increase the tolerance of polydispersity,[12,13] fractionation is necessary for crystallization in highly polydispersed colloids.[14] By fractionation, colloids can crystallize into multiple crystal domains in which the polydispersity is below the terminal polydispersity.[15] Experimental evidences of fractionation have been reported in colloids with size-dependent short-range depletion attractions.[16,17] In these studies, however, direct investigation on the dynamics of fractionation is missing and the underlying mechanism which drives fractionation was not discussed.

A deep understanding of fractionation for self-assembly in multicomponent mixtures has important implication in biology. In living cells, there are a large number of membraneless organelles which are composed of proteins, nucleic acids or other biomolecules. These organelles are widely involved in regulating biological functions and metabolic processes.[18] The formation of membraneless organelles, which are enriched in one or a few components, is a self-assembly process accompanied by segregation or fractionation. Most importantly, the membraneless organelles can form and dissolve, suggesting that the fractionations are reversible. There are an increasing number of studies arguing that the membraneless organelles form as the result of phase separation.[18] However, quantitative evidence is missing.[19] Little is known about the mechanism by which fractionation and phase separation are synchronized.

In this study, to explore the mechanism of fractionation in multicomponent mixtures, crystallization of binary colloids is studied in a two-dimensional (2D) system. As the system is supersaturated in a quasistatic way, crystallization proceeds via a two-step process in which small particles and large particles are fractionated as coexisting crystal and liquid phases respectively. We show that the synchronization of fractionation and phase separation results from the competition between a size-dependent dipolar repulsion and a tunable long-range attraction: tuning the strength of attraction can selectively supersaturate one component for crystallization while maintain other components unsaturated. The observations here serve a direct experimental demonstration of fractionation, and reveal one reversible mechanism by which fractionation is coupled with phase separation. This offers new light on understanding of the formation of membraneless organelles in living cells.

**Methods**

Polystyrene particles (Duke standards, std.dev.< 1.5%) of two distinct diameters of $d$=1.3 $\mu m$ and $d$= 2.0 $\mu m$ are mixed and dispersed in deionized water. The solid volume fractions of small particles and large particles are designed to be 0.10% and 0.38% respectively such that the number ratio of small particles and large particles is around 1:1. The colloidal suspension is sealed between two conducting glass plates coated with indium tin oxide (ITO) (Fig.1a). As an alternating electric field (AEF) is applied, the particles are transferred onto the surface of electrodes by the fluid flow and form a two-dimensional (2D) system (Fig.1b). Depending on the field strength and the frequency, phases including crystal and gas form (Fig.1c). The phase behavior is dominated by the competition between a dipole-dipole repulsion and an electrohydrodynamic-flow-induced attraction.[6, 20] The repulsion is mainly determined by the field strength, and the attraction is sensitive to frequency. Decreasing frequency can enhance the attraction and trigger crystallization. The system is equilibrated first at a stable 2D gas state (Fig. 1b) and then is supersaturated for crystallization by decreasing frequency. The dynamic process is followed and recorded by a high speed CMOS camera. The positions of particles are located and tracked with IDL routines.[21]

**Results**

Starting from the equilibrium gas state (f=4.0kHz), the system is directly quenched to the crystal region of f=1.0kHz. In this study, the peak-to-peak voltage $V_{pp}$ is fixed at 4.0V which corresponds to a peak field strength of $E_p$=18.2V/mm. The fast and deep quenching gives rise to a homogeneous gel network in which large particles and small particles distribute uniformly (Fig. 1d). The first peak of radial distribution function (RDF) represents the average center-to-center distance between two nearest neighbored particles. Distinct from monodisperse systems, the first peak of RDF of the gel splits into one main peak and two sub-peaks (Fig. 1f). To understand the splitting, the RDFs of small particles and large particles in the gel are plotted respectively for reference. It shows that the sub-peak on the left of the main peak corresponds to the mean center-to-center distance between two neighbored small particles (SS) and the right sub-peak stands for the mean distance between two neighbored large particles (LL). It follows that the main peak in between the sub-peaks represents the mean center-to-center distance between a small particle and a large particle (SL). Since the height of the main peak is significantly larger than that of sub-peaks, the particles in the gel are mainly surrounded by distinct particles, suggesting that no fractionation occurs during the gelation. This is reasonable considering that the system is quenched sharply and long-distance diffusion for fractionation is dynamically frozen.

To explore the possible route for fractionation, starting from the homogeneous gas state, the system is quenched in a quasistatic multi-step way. In each step, the frequency is decreased by a small quantity and maintained for a few minutes such that the system has enough time to relax for equilibrium. Distinct from the fast quench, the multi-step quenching produces an equilibrium state consisting of crystalline domains which are composed of identical particles. In the final equilibrium state, small particles and large particles are fractionated as distinct crystalline domains. The resulting RDF consists of three sub-peaks in the first peak (Fig.1g). Distinct from the homogeneous gel structure, the sub-peak of SS becomes the highest. Nevertheless, the height difference between three sub-peaks is not significant, suggesting that the neighboring in the crystalline domains is dominated by SS and LL. It follows that in the multi-step quenching process, particles are separated according to the size as Fig.1e shows. Distinct from previous studies,[16, 17] size- and time- dependent short-range depletion attraction is not present in our study. The fractionation here is induced by a long-range attraction rather than by long-range repulsive forces. [13]

To reveal the underlying mechanism of fractionation, the dynamic process of fractionation is followed and studied. As the system is quenched into the crystal-gas coexisting region but not far away from the crystal-gas boundary (f=3.0kHz), small particles begin to crystallize first while large particles remain in the gas phase. In the starting homogeneous gas (f=4.0 kHz), the first peak of RDF is dominated by the SL signal (green curve in Fig.2d) while the signals for SS and LL are not significant. As the small particles begin to crystallize, the signal of SS emerges as a sub-peak on the left of the main peak of SL (blue curve in Fig. 2d). As the frequency goes down to 2.5 kHz, all small particles crystallize and segregate from the gas phase of large particles (Fig. 2a). As a result, small particles and large particles are segregated as coexisting crystal and gas phases respectively.

Fractionation is accomplished via phase separation. Correspondingly, the signal of SS in the first peak of RDF becomes the main peak while the signal of SL decays as a sub-peak (red curve in Fig. 2d). Large particles began to condense and crystallize around the existing crystallites of small particles as the frequency decreases to 2.0 kHz (Fig.2b). The corresponding signal of SL in RDF (black curve in Fig. 2d) becomes stronger. As the frequency goes down to 1.0 kHz, all large particles condense into the crystal phase, and the signals of SS, SL and LL in the first peak of RDF become comparable in strength. In the final crystal phase, two types of crystalline domains characterized by distinct lattice constants, namely large constant crystals (LC) and small constant crystals (SC), coexist (Fig. 2c). The epitaxial growth of LC around SC results in a large number of boundary particles, which contributes to the signal of SL.

The observed fractionation results from the two-step crystallization in which small particles and large particles nucleate sequentially. It follows that the degree of supersaturation for crystallization in the mixture is size-dependent. We suggest that the size-dependent supersaturation originates from the competition between a size-dependent repulsion and a flow-induced tunable attraction. The particles dispersed in the solution become polarized by the electric field such that they interact with a dipole-dipole repulsion (Fig. 3a) $F_{dip}$ which scales with the particle radius $a$ and the field strength $E$ by $F_{dip} \sim E^2 a^6 C^2$.[22] Here, $C$ is the polarizability of the particle which scales with the frequency $f$ by $C \sim 1/(1+f^2 \tau_{MW}^2)$ where $\tau_{MW}$ represents the relaxation time of surface charges and is in the range of $10^{-9} \sim 10^{-6}$ s. It follows that the dipolar repulsion is very sensitive to the size of particles but depends on frequency weakly. According to this result, the dipolar repulsion between large particles in this study is 13 times stronger than that between smaller particles. On the other hand, the presence of charged particles near electrodes distorts the local electric field and produces transverse flows due to the electrohydrodynamic (EHD) mechanism. The fluid flows bring particles toward each other and produce a long-range attraction $F_{EHD}$ between them.[23] The resulting attraction scales with the field strength and the frequency by $F_{EHD} \sim E^2/f$. Therefore, decreasing frequency can enhance the attraction while its effect on the dipolar repulsion can be neglected relatively.

The competition between the repulsion $F_{dip}$ and the attraction $F_{EHD}$ leads to the size-dependent supersaturation for crystallization, and thus the fractionation via phase separation. The mechanism is that the strength of dipolar repulsion $F_{dip}$ in the binary system is size-dependent as $F_{dip}^{SS} < F_{dip}^{SL} < F_{dip}^{LL}$ (Fig.3). The dipolar repulsion between small particles $F_{dip}^{SS}$ is the weakest. In the initial gas phase, no aggregation occurs, suggesting that the attraction is not strong enough to overcome the weakest dipolar repulsions for aggregation, $F_{EHD} < F_{dip}^{SS}$. As frequency decreases, the attraction increases. As $F_{EHD}$ goes up to the range of $F_{dip}^{SS} < F_{EHD} < F_{dip}^{SL}$, small particles become supersaturated for crystallization. Further increase of $F_{EHD}$ in the limit of $F_{EHD} < F_{dip}^{SL}$ brings more and more small particles into the crystalline phase while large particles are maintained in the gas phase (Fig.2a). As $F_{EHD}$ goes up to the range of $F_{dip}^{SL} < F_{EHD} < F_{dip}^{LL}$, bonding between large particle and small particles becomes possible. As a result, a layer of large particles forms around the crystallites of small particles (Fig. 2b). Crystallization of the rest large particles becomes possible only as $F_{EHD} > F_{dip}^{LL}$.

In contrast to the two-step crystallization, the sharp increase of $F_{EHD}$ from $F_{EHD} < F_{dip}^{SS}$ to $F_{EHD} > F_{dip}^{LL}$ gives rise to gelation as shown in Fig. 1d. Nevertheless, the strength of $F_{EHD}$ can be reversibly tuned by frequency. If the fractionation in the two-step crystallization actually arises from the competition between size-dependent repulsion and the frequency-dependent attraction, we can selectively remove the large particles from the gel by decreasing $F_{EHD}$ from $F_{EHD} > F_{dip}^{LL}$ to $F_{dip}^{SS} < F_{EHD} < F_{dip}^{SL}$ such that large particles cannot be contained in the gel due to the repulsion. As a result, the large particles will be repelled out from the gel and the survived aggregation of small particles can rearrange into crystalline configuration. Further decease of $F_{EHD}$ to $F_{EHD} < F_{dip}^{SS}$ leads to the complete melting of gel. Therefore, a two-step melting of gel may also result in fractionation. To examine the scenario of two-step melting, observation is conducted with the gel. Starting from the gel (Fig. 1d), $F_{EHD}$ is decreased gradually by increasing frequency. As the frequency goes up to 2.0 kHz where $F_{dip}^{SL} < F_{EHD} < F_{dip}^{LL}$ as Fig.2b indicates, large particles located at the edge of gel become free while the large particles inside the gel are still bonded due to the attraction from surrounding small particles (Fig.4a). As a reflection of the part melting, the first peak of RDF in the gel becomes dominated by the signal of SL (black curve in Fig.4d). As frequency goes up to 2.3 kHz, most large particles are rejected from the gel into the gas phase as expected (Fig.4b). Correspondingly, the peak of LL in RDF is separated from the first peak, and the signal of SS gets enhanced (red curve in Fig.4d). There is still a strong signal of SL in the first peak. We suggest that the remained signal of SL arises from the interface between the crystal phase of small particles and the gas phase of large particles. Further increase of frequency gives rise to a complete release of large particles, and the survived aggregations of small particles adapt a crystalline configuration (Fig.4c). As a result, large particles and small particles are separated as two coexisting phases. The melting-induced fractionation confirms the understanding that the phase behavior of the binary systems is dominated by the competition between a size-dependent repulsion and a frequency-dependent attraction.

The above observations offer a direct experimental demonstration of the dynamics of fractionation, and have important implication in biology. In living cells, there are many membraneless organelles which carry out biological functions. These organelles are highly dynamic and exhibit liquid-like behavior, such as dripping, fusion, wetting and reversible deforming.[24] The solution in living cell contains thousands of different protein and RNA molecules. To form functional organelles which are enriched of one or a few components, fractionation is prerequisite. Moreover, the membraneless organelles can form and disappear as needed, suggesting that the fractionation processes have to be reversible. There are an increasing number of studies suggesting that the formation of membraneless organelles results from phase separations.[25-27] This conceptual understanding is being extensively employed to account for the observations of the compartmentalization in living cells.[26, 28] Nevertherless, McSwiggen *et al.*[19] found that most observations about the phase separations in living cells are phenomenological instead of quantitative measurement. Consequently, little is known about the mechanism by which fractionation is synchronized with the phase separation for the reversible formation of membraneless organelles.

Here, in this study, we show that because of the heterogeneous interactions, the degree of supersaturation for phase separation in multicomponent systems is distinct for different components at a given condition. As a result, phase separation proceeds with component fractionation. This observation demonstrates a mechanism by which fractionation is intrinsically synchronized with phase separation. This mechanism can be applied to understand the reversible formation of membraneless organelles in living cells. The interactions between intracellular molecules are mediated by multivalent domains or intrinsically disordered regions (IDRs) which are ubiquitous in proteins and RNA molecules.[25] The strength distribution of multivalent interactions in living cells is as wide as a few $k_B T$.[29] In consistence, simulations found that phase separation becomes common in multicomponent mixtures as the distribution range of interaction is up to a few $k_B T$.[4] Fractionation is more likely to happen as the distribution of interactions becomes broader.[30, 31] In living cells, the compartmentalization for membraneless organelles is generally controlled by changing pH or local ionic strength which can reversibly modify the multivalent interactions, and thus selectively trigger the phase separation of one or a few components.[32] This is in consistent with our observations.

**Conclusions**

In summary, the observations in this study demonstrate a robust mechanism by which fractionation is intrinsically synchronized with phase separation. This mechanism has wide implication in understanding the reversible formation of membraneless organelles in living cells and offers a deeper understanding on how the reversible fractionation and phase separation can be turned on and off by elegantly tuning the attraction between components.

**Acknowledgments**

T. H. Zhang acknowledges the financial support from the National Natural Science Foundation of China under Grant No. 11674235, 11635002 and 11974255.

**Figures**

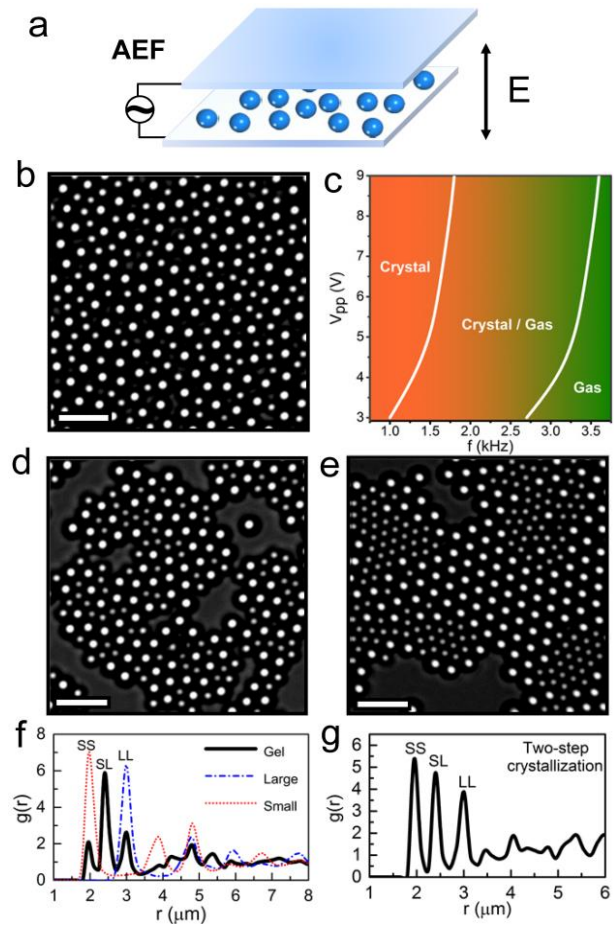

**Fig.1** Gelation and Fractionation. a) Experimental setup. The colloidal system is controlled by an alternating electric field. b) Equilibrium gas before supersaturation for crystallization at f=4.0kHz. c) Phase diagram of the 2D system. d) Gel at f=1.0kHz resulting from a sharp quenching. e) Crystalline structures after fractionation at f=1.0kHz resulting from a multi-step quenching process. f) Pair radial distribution of the gel in d. g) Pair radial distribution of the crystalline structures in e.

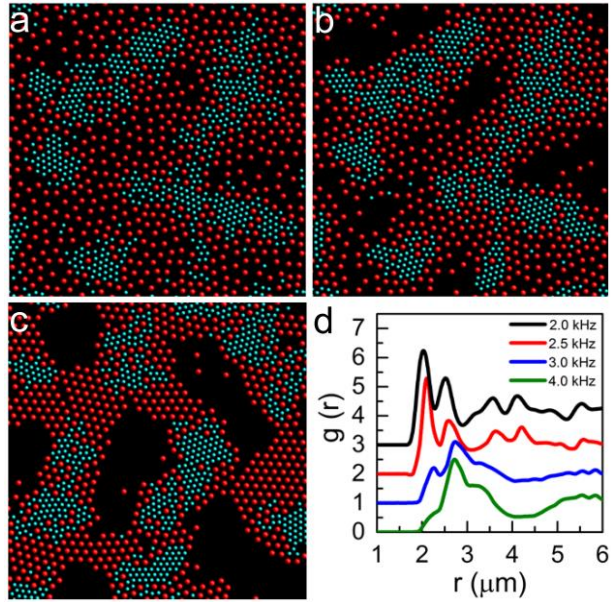

**Fig. 2** Process of fractionation in the two-step crystallization. a) Coexisting crystal and gas phases at f=2.5kHz. b) Part crystallization of large particles at f=2.0kHz. c) Multi-crystal phase at f=1.0kHz. d) Evolution of pair radial distribution during the fractionation. Red sphere: Large particle. Blue sphere: small particle.

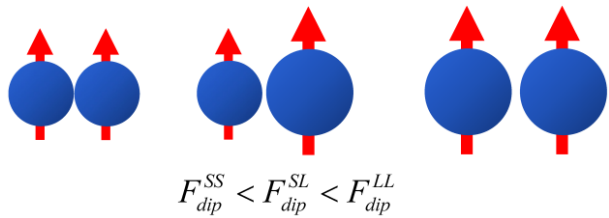

**Fig. 3** Illustration of the polar repulsion between colloidal particles.

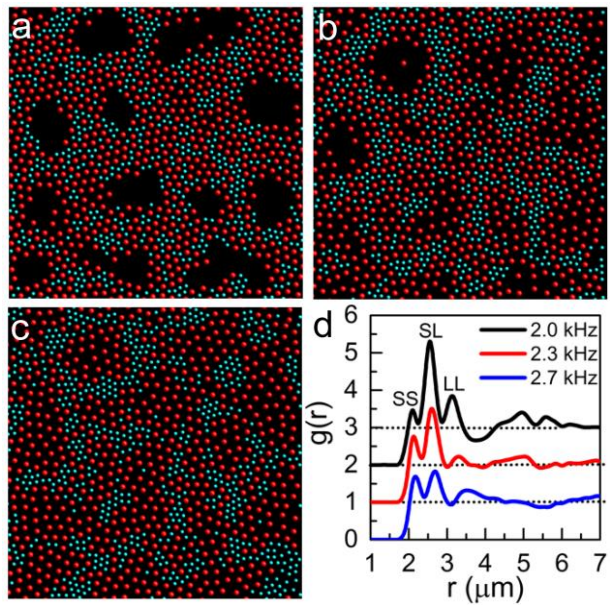

**Fig. 4** Fractionation in two-step melting. a) Meting gel at f=2.0kHz. b) Partly melt gel at f=2.3kHz. c) Coexisting crystal and gas phases resulting from the melting of gel at f=2.7kHz. d) Evolution of pair radial distribution during the melting. Red sphere: Large particle. Blue sphere: small particle.